\documentstyle[prl,twocolumn,aps,epsf]{revtex}
\begin{document}
\draft
\twocolumn[\hsize\textwidth\columnwidth\hsize\csname@twocolumnfalse%
\endcsname

\title{Hole Pairs in the Two-Dimensional Hubbard Model}
\author{E. Louis}
\address{Departamento de F{\'\i}sica Aplicada,
Universidad de Alicante, Apartado 99, E-03080 Alicante, Spain.}

\author{F. Guinea, M. P. L\'opez Sancho, J. A. Verg\'es}
\address{{I}nstituto de Ciencia de Materiales de Madrid,
CSIC, Cantoblanco, E-28049 Madrid, Spain.}

\date{\today}

\maketitle

\begin{abstract}
The interactions between holes in the Hubbard model, in the low
density, intermediate to strong coupling limit, are investigated.
Dressed spin polarons in neighboring sites have an increased kinetic
energy and an enhanced hopping rate. Both effects are of the
order of the hopping integral and lead to an effective attraction
at intermediate couplings. Our results are derived by systematically
improving mean field calculations. The method can also be used to
derive known properties of isolated spin polarons.
\end{abstract}

]
\narrowtext
The nature of the low energy excitations in the Hubbard model has
attracted a great deal of attention. Close to half filling,
a large amount of work suggests the existence of spin polarons,
made of dressed holes, which propagate within a given sublattice
with kinetic energy of the order of
$J = \frac{4 t^2}{U}$\cite{BSW94,PLH95}, where $t$ is the hopping
integral and $U$ the on site Coulomb repulsion.
These results are consistent with similar calculations
in the strong coupling,
low doping limit of the Hubbard model, the $t-J$ 
model\cite{DNB94,LG95,DNHMRD97}.
There is also evidence for an effective attraction between these
spin polarons\cite{FO90,BM93,D94,KA97,ZC97}.

In the present work, we analyze the dynamics of spin polarons and the
interactions between them by means of a systematic expansion
around mean field calculations of the Hubbard model on the bipartite
square lattice. 
Two spin polarons in neighboring sites experience an increase
in their internal kinetic energy, due to the overlap of the
charge cloud. This repulsion is of the order of $t$. 
In addition, a polaron reduces the obstacles
for the diffussion of another, leading to an assisted hopping
term which is also of the order of $t$. The combination of these
effects is an attractive interaction at intermediate values of
$U/t$.

Use of the Unrestricted Hartree Fock (UHF) approximation in finite clusters
provides a first order approximation to the spin polaron near half
filling. As discussed elsewhere, this approximation
describes well the undoped, insulating
state at half filling \cite{VL91}. A realistic picture of
the spin wave excitations is obtained by adding
harmonic fluctuations by means of the time dependent Hartree Fock
approximation (RPA)\cite{GLV92}. At intermediate and large
values of $U/t$, the most stable HF solution with a single hole is
a spin polaron\cite{VL91}. This solution is replaced by a fully ferromagnetic
one at sufficiently large values of $U/t$ \cite{LC93a}.
Approximately half of the charge of the hole
is located at a given site in the spin polaron solution.
The spin at that site is small and it is reversed with
respect to the antiferromagnetic background. The remaining charge
is concentrated in the four neighboring sites.
A number of alternative derivations lead to a similar picture
of this small spin bag\cite{Hi87,KS90,DS90,A94}.
A similar solution is expected
to exist in the $t-J$ model.

As usual in mean field theories, the spin polaron solution described
above breaks symmetries which must be restored by quantum fluctuations.
In particular, it breaks spin symmetry and translational
invariance. Spin isotropy must exist in finite clusters. However, it
is spontaneously broken in the thermodynamic limit, due to the
presence of the antiferromagnetic background. Hence, we do not expect
that the lack of spin invariance is a serious drawback of the
Hartree Fock solutions. In any case, spin isotropy can be restored, starting
from the mean field wavefunction, by projecting out the components
which do not have a predetermined spin.
Results obtained for small clusters \cite{LC93b,LG92} show a slight
improvement of the energy,
which goes to zero as the cluster size is increased.
On the other hand, translational invariance is expected
to be present in the exact solution of clusters of any size.
Thus, we have improved the mean field results by
hybridizing a given spin polaron solution with all
wavefunctions obtained from it by lattice translations.
This procedure is equivalent to the configuration interaction (CI)
method used in quantum chemistry. 
If the initial mean field
solution is considered as the ^^ ^^ classical" zeroth order
approximation to the exact solution, this scheme can be
described as the inclusion of instanton effects, in which
the spin polarons tunnel between equivalent configurations.
Finally, in addition to the previous corrections, we can 
add the zero point fluctuations around the RPA ground state\cite{GLV92}.
This calculation does not change appreciably the results, although it
is necessary to describe correctly the long ranged magnon cloud around 
the spin polaron\cite{RH97}.

\begin{figure}
\begin{center}
\mbox{\epsfxsize 8cm\epsfbox{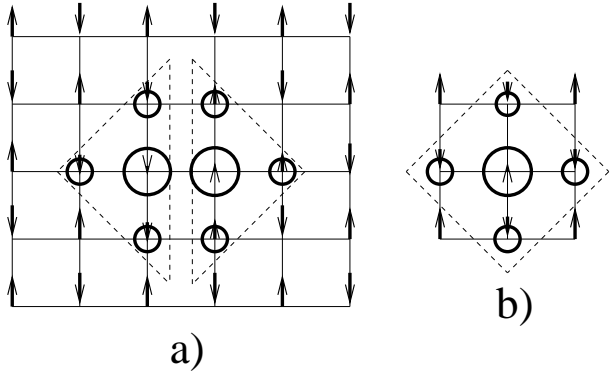}}
\end{center}
\caption{
a) Sketch of one of the bipolaron solutions,
at large values of $U/t$, considered in the text.
Circles denote the local charge, measured from half filling,
and arrows denote the spins.
There are two localized states marked by the dashed line.
For comparison, the single polaron solution is shown in b).}
\label{polaron1}
\end{figure}

A schematic picture of the initial one hole and two holes Hartree Fock
wavefunctions used in this work is shown in Fig. \ref{polaron1}.
They represent the solutions observed at large
values of $U/t$ for the isolated polaron and two spin polarons
on neighboring sites. Actually, charge localization is associated
to the existence of bound states
which split from the top of the lower Hubbard band.

\begin{figure}
\begin{picture}(235,225) (-0,20)
\epsfbox{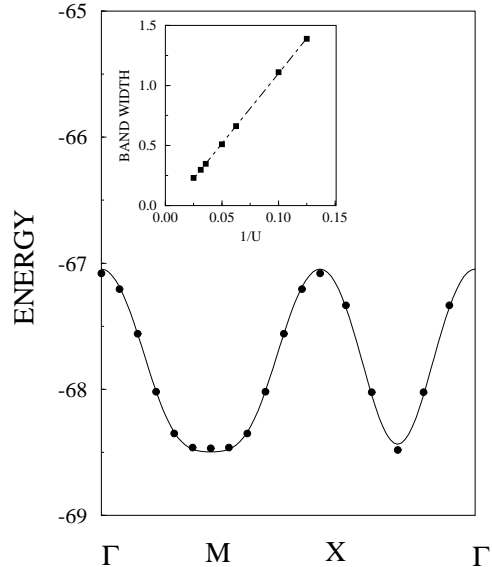}
\end{picture}
\caption{
Quasiparticle band structure for a single hole on $12 \times 12$ clusters
of the square lattice with periodic boundary conditions and $U=8t$
(filled circles). The continuous line corresponds to the fitted
dispersion relation (see text). The inset
shows the bandwidth as a function of $t^2/U$;
the fitted straight line is $-0.022 t + 11.11 t^2/U$.
\label{polaron}}
\end{figure}

Spin polaron wavefunctions localized at different sites are not orthogonal.
Both wavefunctions overlap and non--diagonal matrix elements of
the Hamiltonian need to be taken into account when mixing between
configurations is considered. The procedure has
been described in detail in\cite{LC93b}, along with the
energy improvements (with respect to UHF) introduced by this CI scheme.
Calculations have been carried out on $L \times L$ clusters with
periodic boundary conditions ($L \leq 12$) and $U \ge 8t$ \cite{Uvalues}.
Although larger clusters can be easily reached, no improvement
of the results is achieved due to the short--range character
of the interactions. The calculated dispersion of a
single polaron is shown in Fig. \ref{polaron}.
Because of the antiferromagnetic background,
the band has twice the lattice periodicity.
Exact calculations in finite clusters do not show
this periodicity, as the solutions have a well defined spin
and mix different background textures. As cluster sizes are increased,
however, exact solutions tend to show the extra periodicity
of our results. We interpret it as a manifestation
that spin invariance is broken in the thermodynamic
limit, because of the antiferromagnetic background.
Hence, the lack of this symmetry in our calculations
should not induce spurious effects.
The only overlaps and matrix elements which are not
negligible are those between polaron wavefunctions
located in the same sublattice.
Fig. \ref{polaron} shows the polaron bandwidth
as a function of $U$. It behaves
as $t^2/U$ (the fitted law is given
in the caption of Fig. \ref{polaron}).
Our scheme allows a straightforward explanation of
this scaling.
Without reversing the spin of the whole background,
the polaron can only hop within a given sublattice.
This implies an intermediate virtual hop into a site with
an almost fully localized electron of the opposite spin.
The amplitude of finding a
reversed spin in this new site decays as $t^2/U$ at large $U$.
We find that the polaron band can be very well fitted by
the expression: $ \epsilon_k = \epsilon_0 + 4 t_{11} \cos (
k_x ) \cos ( k_y ) + 2 t_{20} [ \cos ( 2 k_x ) + \cos ( 2 k_y ) ]
+ 4 t_{22} \cos ( 2 k_x ) \cos ( 2 k_y ) +
4 t_{31}[ \cos ( 3 k_x) \cos ( k_y )+ \cos ( k_x ) \cos ( 3 k_y )]$.
For $U = 8 t$, we
get $t_{11} = 0.1899 t$ , $t_{20} = 0.0873 t$, $t_{22} = -0.0136 t$,
and $t_{31} = -0.0087 t$.
All hopping integrals vanish as $t^2/U$
in the large $U$ limit for the reason given above.
Also the energy gain with respect to UHF \cite{LC93b}
behaves in this way. Let us mention, that all
the features reported here are in good agreement with known
results\cite{BSW94,PLH95,DNB94,LG95} for both
the Hubbard and the $t-J$ models.

\begin{figure}
\begin{picture}(235,150) (-20,-30)
\epsfbox{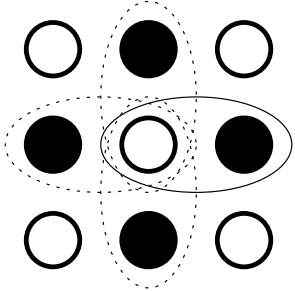}
\end{picture}
\caption{
Sketch of the bipolaron UHF wavefunctions used in this work. Note that the 
four wavefunctions are obtained by successive rotations
of $\pi /2$. The complete basis set is produced by translation of these
wavefunctions through the whole cluster.
\label{bipolaronCI}}
\end{figure}

We now consider solutions with two spin polarons.
The relevant UHF solutions are those with $S_z = 0$.
In order to get some coupling, the centers of the
two spin polarons must be located in different sublattices.
The mean field energy increases as the two polarons
are brought closer, although, for intermediate and large
values of $U$, a locally stable Hartree Fock solution can be
found with two polarons at arbitrary distances.
We have not attempted to do a full CI
analysis of all possible combinations of two holes in a finite
cluster. Instead, we have chosen a given mean field solution
and hybridized it with all others obtained by a lattice translation
or rotation. Clusters of sizes up to $10 \times 10$ were studied
which, as in the case of the polaron, are large enough due to the 
short--range interactions between different configurations. 
The basis used for the two polarons at the shortest distance
is shown in Fig. \ref{bipolaronCI}.
This procedure leads to a set of bands for the two
hole configurations. The number of bands is two or four, depending on
the number of different configurations which can be obtained by
rotations at a given site. Like in the single polaron case,
we obtain a gain in energy, due to the delocalization of the pair.

\begin{figure}
\begin{picture}(235,225) (-20,-10)
\epsfbox{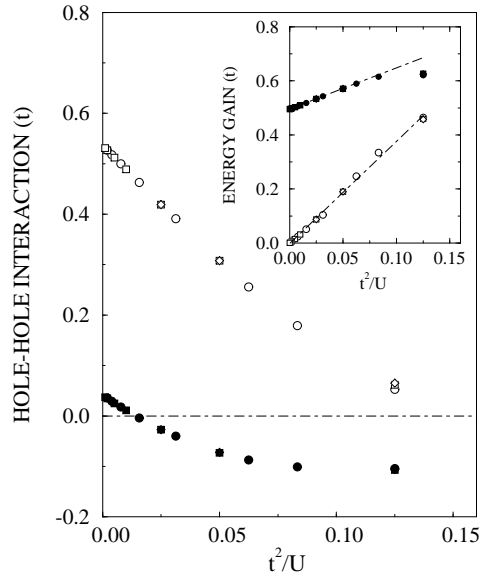}
\end{picture}
\caption{
Comparison of the hole-hole interaction (see main text for the definition)
obtained within UHF (empty symbols) and CI (filled symbols) approximations.
Results correspond to $6 \times 6$, (circles) $8 \times 8$
(squares) and $10 \times 10$ (diamonds) clusters with periodic boundary
conditions.
The inset shows the energy gain due to the inclusion of correlation
effects via CI for both the configuration of holes located in
neighboring positions (filled simbols) and holes that are maximally
separated in the finite size cluster (empty simbols).
The fitted straight lines are:
$0.495 t + 1.53 t^2/U$ (filled symbols) and
$-0.002 t + 3.78 t^2/U$ (empty symbols).
\label{difference}}
\end{figure}

The hole-hole interaction, i.e., the difference
between the energy of a state built up by all configurations with the
two holes at the shortest distance (separated by a vector of the set
\{1,0\}) and the energy of the state having the holes at the largest
distance possible at a given cluster is depicted in Fig. \ref{difference}.
Two holes bind for intermediate values of $U$ \cite{ferro}. 
This happens because the delocalization energy tends to be higher
than the repulsive contribution obtained within mean field.
The local character of the interactions is
illustrated by the almost null dependence of the results
shown in Fig. \ref{difference} on the cluster size. 
The energy gain of the two holes (with respect to UHF) in the
two limiting configurations (at the shortest or largest distance possible)
is given in the inset of Fig. \ref{difference}. Note that, whereas 
in the case of the holes at the largest distance, the gain goes to zero
in the large $U$ limit, as for the isolated polaron, when
the holes are separated by a $\{1,0\}$ vector the gain goes to a
finite value. This result is not surprising, as the arguments given below
suggest, and is in line with the results for the width of the
quasiparticle band. The numerical results for $L$=6, 8 and 10 and $U$
in the range $8t-5000t$ can be fitted by the following straight lines,
$3.965 t + 14.47 t^2/U$ (holes at the shortest distance) 
and $-0.007 t + 10.1 t^2/U$ (holes at the largest distance). 
Thus, total bandwidth of the two bands obtained for holes in neighboring
sites does not vanish in the infinite $U$ limit (as the energy gain reported
in Fig. 4). The internal consistency of
our calculations is shown comparing the large $U$ behavior
of the two holes at the largest distance possible with the corresponding
results obtained for the isolated polaron (compare this fitting with
that given in the captions of Fig. \ref{polaron}) .

Both the
behavior of the quasiparticle bandwidth and that of the energy
gain with respect to UHF of the two holes at the shortest distance
can be understood from arguments
similar to those used for the single polaron. 
The hopping terms that are proportional to $t$ at large $U$
describe the rotation of a pair around the position of one of
the two holes. Each hole is spread between four sites.
In order for a rotation to take place, one hole has to jump
from one of these sites into one of the rotated positions.
This process can always take place without a hole moving into a
almost fully polarized site with the wrong spin. In the single
polaron case, as discussed before, the motion of a hole
involves the inversion of, at least, one almost fully spin polarized
site, in the large $U$ limit. 
As a consequence, the delocalization of polaron pairs on
neighboring sites leads to a 
finite gain in energy, even as $U \rightarrow \infty$, as opposed
to the single polaron case \cite{LC93b} or polaron pairs at the largest
distance (see inset of Fig. \ref{difference}).

The possibility of hole assisted hopping was discussed in\cite{Hi93},
in a different context. It always leads to superconductivity.
In our case, we find a contribution,
in the large $U$ limit, of the type:
\begin{eqnarray}
{\cal H}_{hop} &= &\sum \Delta t {c^{\dag}}_{i,j;s} c_{i,j;s} (
{c^{\dag}}_{i+1,j;{\bar s}} c_{i,j+1;{\bar s}} +  
\nonumber \\ &+ &{c^{\dag}}_{i-1,j;\bar{s}} c_{i+1,j;\bar{s}} + h. c. +{\rm perm})
\label{hopping}
\end{eqnarray}
This term admits the BCS decoupling 
$\Delta t \langle c^{\dag}_{i,j;s}
c^{\dag}_{i+1,j;{\bar s}} \rangle c_{i,j;s} c_{i,j+1;{\bar s}} +
h. c.  + ...$. 
It favors superconductivity with
either $s$ or $d$ wave symmetry, depending on the sign of $\Delta t$.
Since we find $\Delta t > 0$, d wave symmetry follows.

An interesting question is whether the holes would tend to segregate
when more holes are added to the cluster. In order to investigate this 
question, we have calculated the total energies for four holes
centered on a square and two separated bipolarons with holes
at the shortest distance. Two (four)
configurations (plus translations) were included in each case.
The results for 4 holes on a $8 \times 8$ cluster and $U=8t$ are, 
$-34.06t$ (four holes on a square) and $-34.48t$ (two bipolarons). Note
that although the fourfold--polaron has also hopping terms which 
do not vanish in the infinite $U$ limit, they are weaker than in the
bipolaron case. These results indicate that for large
and intermediate $U$ no hole segregation takes place (for small
$U$ see below) and
that the most likely configuration is that of separated bipolarons.

The picture presented above is consistent with other analytical
and numerical studies of hole-hole pairing
in real space\cite{Do96,WS97,RD97,corr} at low
fillings and intermediate to large values of $U/t$.
As $U$ is reduced, the size of the spin polarons increases
and becomes elongated along the diagonals of the square lattice.
The most likely solutions of the Hartree-Fock
calculations are domain walls which separate antiferromagnetic
regions\cite{ZG89,PR89,Su90,VL91}. The breakdown of translational
symmetry associated with these solutions is probably real and not
just an artifact of the Hartree Fock solution, as in the case
discussed previously. Hence, we expect a sharp transition between
a regime of small spin polarons with an effective attraction
and striped phases at low $U$. Note, however, that the
scheme presented here, based on mean field solutions plus
corrections, is equally valid in both cases.

Summarising, we have analyzed the dynamics of spin polarons
and their interactions by systematically improving the
mean field approximation to the Hubbard model. Our scheme
gives an intutive framework in which the appearance of attraction between
holes can be understood.

Financial support from the spanish CICYT (through grants PB96-0875,
PB96-0085, PB95-0069) is gratefully acknowledged.


\begin{references}
\bibitem{Uvalues}
Note that $U=8t$ is widely accepted as the most physically meaningful
value of Coulomb repulsion in these systems (see for instance \cite{WS97}.
\bibitem{ferro}
Note that at large $U$ the ferromagnetic solution sets in.
In Ref. \cite{LC93a} it was shown that the energy of two holes in
a ferromagnetic $L \times L$ cluster could be fitted by
$E/t=-7.942+31.02/L^2$.
Comparing the present numerical results with this energy gives a transition
to the ferromagnetic solution approximately at $U/t=32$, for the cluster
sizes investigated here.
\bibitem{corr}
Note that, due to the delocalization of the polarons,
 the probability of finding the two holes in the same sublattice
is of the same order as that of finding them in different sublattices
(see fig.\ref{polaron1}). This is in agreement with\cite{WS97} (see
also\cite{RD97}).
\end{references}
\end{document}